\documentclass{article}
\usepackage{spconfa4,amsmath,graphicx}
\usepackage{todonotes}
\usepackage[nice]{nicefrac}
\usepackage{siunitx}
\usepackage{booktabs}
\usepackage{multirow}
\usepackage{float}

\makeatletter
\def\section{\@startsection{section}{1}{\z@}%
  {1.8ex plus 0.7ex minus 0.3ex}%
  {0.9ex plus 0.3ex}%
  {\normalfont\normalsize\bfseries\centering}}
\def\subsection{\@startsection{subsection}{2}{\z@}%
  {1.3ex plus 0.5ex minus 0.2ex}%
  {0.7ex plus 0.2ex}%
  {\normalfont\normalsize\bfseries}}
\def\subsubsection{\@startsection{subsubsection}{3}{\z@}%
  {1.0ex plus 0.4ex minus 0.2ex}%
  {0.5ex plus 0.2ex}%
  {\normalfont\normalsize\itshape}}
\makeatother

\title{Optimized Compressor Parameterization\\for Produced Music in Hearing Aids}
\name{Robin Holland\textsuperscript{1,2}, Christoph Hoog Antink\textsuperscript{1}, Henning Puder\textsuperscript{1,2}}
\address{
\textsuperscript{1}Technische Universität Darmstadt, Darmstadt, Germany\\
\textsuperscript{2}WSA, Erlangen, Germany
}
\begin{document}
\raggedbottom
%
\maketitle
\begin{abstract} 
Hearing aid compression is typically designed for speech, yet music demands different trade-offs due to its narrower dynamic range (DR) and impulsive transients. This paper investigates whether music fitting benefits from being signal-aware, both in mapping compression demand across signal–audiogram conditions and in tuning its parameters to these combinations rather than inheriting them from speech-derived prescriptions. Compression demand is characterized across eight audiometric profiles, three residual DR severities, and three produced music signals under linear gain. Signal dynamics emerge as a first-order factor: production-compressed music fits the residual DR of the mild-to-moderate profiles evaluated, while dynamically rich recordings challenge even moderate hearing loss. Per-signal optimization of single- and multi-band wide dynamic range compression is then applied to two profiles with the highest compression demand. This optimization improves the Hearing Aid Audio Quality Index over gain-optimized NAL-NL2 baselines while also improving audibility, indicating substantial headroom for music-specific fitting.
\end{abstract}
\section{Introduction}
\label{sec:intro}

Music listening is an important use case for hearing aids, yet hearing aid design and evaluation remain focused on speech intelligibility~\cite{Zakis2016}. Produced (i.e., mixed and mastered) music differs from speech in ways relevant to hearing aid processing: it typically exhibits a narrower dynamic range (DR)~\cite{kirchbergerDynamicRangeMusic2016a} and is often listened to at higher levels~\cite{croghanChosenListeningLevels2016, serpanosEffectSoundPreference2025}. These properties place demands on wide dynamic range compression (WDRC) that speech-oriented designs are not optimized to meet.

Commercial hearing aids often offer dedicated music programs, typically with slower compression and reduced noise reduction~\cite{Zakis2016,mooreEffectsSoundInducedHearing2016}. However, evidence for their effectiveness is inconclusive~\cite{greasleyMusicListeningHearing2020a, greasleyUsingHearingAids2026a}. Objective-function-driven optimization of insertion gains and compression thresholds has been explored for speech intelligibility~\cite{goncalvesbrazOPRARSHearingAidFitting2022}, but direct optimization of compressor parameterization for music remains unexplored.

A key factor in compression demand may be the listener's residual DR, the range between hearing threshold (HTL) and uncomfortable loudness level (UCL). Sensorineural hearing loss (HL) often reduces this range through loudness recruitment and, with hyperacusis, through lowered UCLs~\cite{mooreCochlearHearingLoss2007a}. Hearing aid users generally prefer reduced or no compression for music~\cite{kirchbergerDynamicRangeMusic2016a, croghanMusicPreferencesHearing2014a}, but whether this is possible without sacrificing audibility likely depends on residual DR. As hearing loss becomes more severe, compression becomes more strongly indicated, yet also more detrimental to music quality~\cite{kirchbergerDynamicRangeMusic2016a}. How compression demand varies across audiogram shape and residual DR severity has not been systematically characterized, and commercially released recordings already undergo multi-stage compression~\cite{kirchbergerDynamicRangeMusic2016a}, potentially reducing the need for further processing.

This paper addresses two questions: how compression demand varies across HL profile, residual DR, and music signal dynamics; and whether per-signal parameter optimization improves objective compression quality beyond fixed clinical fittings. The contributions are: (1)~a continuous characterization of compression demand across eight HL profiles with three residual DR classes each, showing that signal dynamics is a first-order factor comparable to residual DR severity; and (2)~per-signal optimization of 1-, 7-, and 20-channel WDRC for two audiograms with strong compression demand, revealing a quality--audibility trade-off shaped by audiogram and signal dynamics, with multi-band architectures improving HAAQI relative to gain-optimized NAL-NL2 baselines~\cite{keidserNALNL2PrescriptionProcedure2011}.
 
 
\section{Data}
\label{sec:data}
\subsection{Music Signal Selection}
Music signals are drawn from the Free Music Archive (FMA)~\cite{defferrardFMADatasetMusic2017} through a filtering procedure aimed at typical music with broad listener appeal: genre filtering to retain standard production practices, ranking by artist popularity, and manual inspection to confirm mainstream production style. The remaining tracks are characterized by two dynamic features: a \textit{spectral macrodynamic measure} (energy-weighted spread across \nicefrac{1}{3}-octave bands) and a \textit{transient microdynamic measure} (broadband peak-to-loudness ratio). Their rank-normalized mean defines a composite dynamism score. From this pool, three signals are selected to represent \textit{Tight}, \textit{Normal}, and \textit{Dynamic} dynamism. The final 20\,s of each clip are used for all evaluations in this paper.

\subsection{Audiogram Dataset}
Audiograms are drawn from the Oldenburg Hearing Health Record (OHHR)~\cite{jafriOldenburgHearingHealth2025}, a clinical dataset with individually measured pure-tone thresholds (up to 4\,kHz) and UCLs. UCLs at unmeasured frequencies are estimated via the Pascoe model~\cite{pascoedavidpedroClinicalMeasurementsAuditory1988}. Each ear is independently classified against the standard audiometric profiles N1--N7 and S1--S3~\cite{bisgaardStandardAudiogramsIEC2010a} by minimizing root-mean-square (RMS) error; N6 and N7 are excluded due to insufficient data. Within each remaining class, ears are stratified into three residual DR groups: \textit{High}, \textit{Moderate}, and \textit{Limited}, by percentile rank of the mean UCL-to-threshold difference. The representative audiogram for each group is the medoid in the joint threshold--UCL space.
 
\section{Analysis}
\label{sec:analysis}

\subsection{Compression Effectiveness Metric}
\label{sec:percentile_effectiveness}

The compression effectiveness (CE) metric quantifies whether the amplified signal fits inside the listener's usable hearing range in each frequency band, e.g. does any part exceed the UCL (risking discomfort), and does any part fall below the threshold (lost to inaudibility). Because music contains both sustained passages and brief transient peaks, the analysis operates at two timescales: a slow \textit{envelope path} (\SI{400}{\milli\second} window, \SI{100}{\milli\second} hop) capturing sustained levels, and a fast \textit{transient path} (\SI{20}{\milli\second} window, \SI{5}{\milli\second} hop) capturing impulsive peaks that the longer window attenuates.

For each \nicefrac{1}{3}-octave band $b$ (center frequencies $b_c=[100,..., 10000]$\,Hz resulting in $B=21$ bands), both paths yield 101-point percentile vectors of short-time RMS levels in dB\,SPL at the eardrum: $\mathbf{P}_b^{\mathrm{env}}$ (envelope) and $\mathbf{P}_b^{\mathrm{tr}}$ (transient), where $P_b(i)$ generically denotes the $i$-th integer percentile ($i = 0,\ldots,100$). Let $T_b$ and $U_b$ denote the HTL and UCL in band $b$, with residual DR $\Delta_b^{\mathrm{res}} = U_b - T_b$.

The upper-bound effectiveness (UE) quantifies the absence of UCL overshoot, and the lower-bound effectiveness (LE) quantifies the absence of threshold undershoot:

{\ninept
\begin{align}
  UE_b &= 1 - \frac{1}{101\,\Delta_b^{\mathrm{res}}}
              \sum_{i=0}^{100} \max\!\bigl(P_b(i) - U_b,\; 0\bigr),
  \label{eq:UE}
\end{align}
\begin{align}
  LE_b &= 1 - \frac{1}{101\,\Delta_b^{\mathrm{res}}}
              \sum_{i=0}^{100} \max\!\bigl(T_b - P_b(i),\; 0\bigr).
  \label{eq:LE}
\end{align}
}
Both scores are clipped to $[0,1]$. Three global ($\mathrm{G}$) scores are derived by averaging across bands:
$\mathrm{GUE} = \frac{1}{B}\sum_{b=1}^{B} UE_b^{\mathrm{env}}$ and
$\mathrm{GLE} = \frac{1}{B}\sum_{b=1}^{B} LE_b^{\mathrm{env}}$ capture sustained UCL compliance and audibility;
$\mathrm{TGUE} = \frac{1}{B}\sum_{b=1}^{B} UE_b^{\mathrm{tr}}$ captures peak UCL compliance.

\subsection{Compression Demand Under Linear Gain}
\label{sec:compression_necessity}

To determine where compression is needed, NAL-R linear prescriptive gain~\cite{byrneNationalAcousticLaboratories1986} is applied to each signal for every audiogram--residual-DR combination. A broadband gain offset is optimized per condition to isolate compression demand from suboptimal broadband level. Conditions where GUE or GLE fall substantially below 1.0 are candidates for WDRC.

{
\begin{figure*}[t]
  \includegraphics[width=\textwidth]{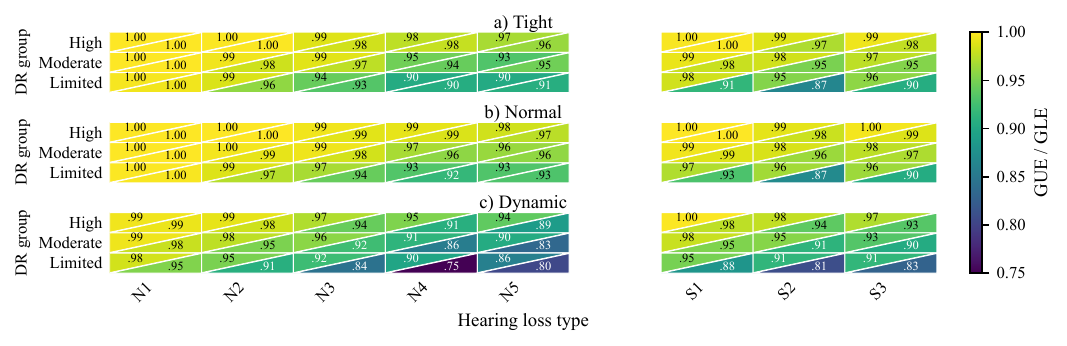}
  \caption{Compression effectiveness (CE) under gain-optimized NAL-R (GR) for three signals: a) Tight, b) Normal, c) Dynamic. Upper triangle: GUE; lower triangle: GLE. HL profile on x-axis, residual DR group on y-axis.}
  \label{fig:heatmap_pre_analysis}
\end{figure*}
}

\subsection{Compressor Parameter Optimization}
\label{sec:param_opt}

Based on the compression demand analysis, two audiometric profiles with limited residual DR are selected as cases clearly requiring compression (see Section~\ref{sec:results}). For these, per signal--audiogram optimization is performed using Optuna~\cite{akibaOptunaNextgenerationHyperparameter2019} with the CMA-ES sampler~\cite{loshchilovCMAESHyperparameterOptimization2016}. Three WDRC configurations are evaluated: \textit{WDRC-1}, \textit{WDRC-7}, and \textit{WDRC-20}, with 1, 7, and 20 log-spaced compression channels, respectively.

Each compression channel uses a branching peak envelope follower with attack and release time constants of 15\,ms and 500\,ms. The static compression characteristic is piecewise linear with hard knees, defined by two thresholds $T_1 < T_2$ and two compression ratios $R_1$ (applied between $T_1$ and $T_2$) and $R_2 > R_1$ (applied above $T_2$). All configurations are initialized from NAL-NL2 targets: the insertion-gain curve and T1 are fitted to match NAL-NL2 at reference input levels and held fixed during optimization. Freeing $T_1$ alongside the compression ratios would yield a non-identifiable parameter space, as different threshold--ratio combinations can produce equivalent input--output mappings. Attack and release times are likewise fixed, so that the optimization targets the static compression characteristic alone.

Per channel, three parameters are free: $T_2$, $R_1$, and $R_2$. A broadband gain offset is applied additively to all insertion gains, allowing each configuration to operate at its optimal overall presentation level so that comparisons reflect compression behavior rather than suboptimal broadband gain. For WDRC-7 and WDRC-20, parameters are optimized at five log-spaced anchor channels and interpolated linearly to the remaining channels, yielding 16 free parameters. WDRC-1 has four free parameters. WDRC-1 receives 300 CMA-ES iterations; WDRC-7 and WDRC-20 each receive 1200.

Objective music quality is assessed using HAAQI~\cite{katesHearingAidAudioQuality2016a}, evaluated against a NAL-R linear-gain reference. The hearing aid insertion gain stage is bypassed prior to scoring, so that HAAQI reflects compression-induced distortion rather than spectral differences from the fitting target. The composite objective balances quality, audibility, and UCL compliance:

{\ninept
\begin{multline*}
    J(\theta) = w_q \cdot \mathrm{HAAQI}(\theta)
              + w_{\mathrm{gle}} \cdot \mathrm{GLE}(\theta)\\
              + w_{\mathrm{tgue}} \cdot \mathrm{TGUE}(\theta)
              - P(\theta)
\end{multline*}
}

with $w_q = 0.4$, $w_{\mathrm{gle}} = 0.4$, and $w_{\mathrm{tgue}} = 0.2$. The weights are fixed a priori rather than optimized; they encode the design intent that perceptual quality and sustained-level compliance are treated as equally important primary objectives. Transient compliance receives a lower weight because brief peak excursions are perceptually better tolerated than sustained ones~\cite{zwislockiTemporalSummationLoudness1969, fortunetoddDurationCompressionAided2000}. The GUE is enforced through the exponential penalty:

{\ninept
\begin{equation*}
    P(\theta) =
    \begin{cases}
        2\,\bigl(\exp\!\bigl(5\,(1 - \mathrm{GUE}(\theta))\bigr) - 1\bigr), & \text{if } \mathrm{GUE}(\theta) < 1 \\
        0, & \text{otherwise}
    \end{cases}
\end{equation*}
}

The objective is evaluated at 55 and 90\,dB\,SPL per trial and averaged. These levels represent the endpoints of a relevant music listening range and stress the conditions where audibility and quality degradation are most critical, respectively. Optimized models are compared against NAL-NL2 (G2) and NAL-R (GR) target baselines. All baselines are optimized with a broadband gain offset.
 
\section{Results \& Discussion}
\label{sec:results}
\subsection{Compression Demand Under Linear Gain}

Fig.~\ref{fig:heatmap_pre_analysis} maps the CE of gain-optimized NAL-R (GR) across HL profiles, residual DR groups, and signals. For mild HL with high residual DR (e.g., N1, N2, S1), both GUE and GLE remain above 0.98 regardless of signal type. CE degrades monotonically with increasing HL severity and decreasing residual DR, reaching GUE\,=\,0.86 and GLE\,=\,0.75 for the worst cases (N4, N5 with limited DR, Dynamic signal).

Signal dynamics are a primary factor: for a given audiogram, GLE spread across signals can reach 0.15, comparable to the effect of residual DR severity. GUE varies by only 0.04, because production-chain mastering likely compresses peaks similarly across signals while leaving quiet-passage levels genre-dependent. Sloping HL (S1--S3) show systematically lower GLE than flat HL of comparable severity. For the profiles and signals evaluated, linear gain is sufficient in most conditions. Profiles N4 and S2 with limited residual DR are selected for optimization as cases where compression is clearly necessary (Fig.~\ref{fig:audiograms}), both exhibiting GUE\,$<$\,0.93 and GLE\,$<$\,0.87 for at least one signal. It should be noted that these calculated metrics represent numerical trends that require validation through subjective testing to establish their perceptual significance and clinical utility.

{
\begin{figure}[h]
    \centering
    \includegraphics[width=\linewidth-2cm]{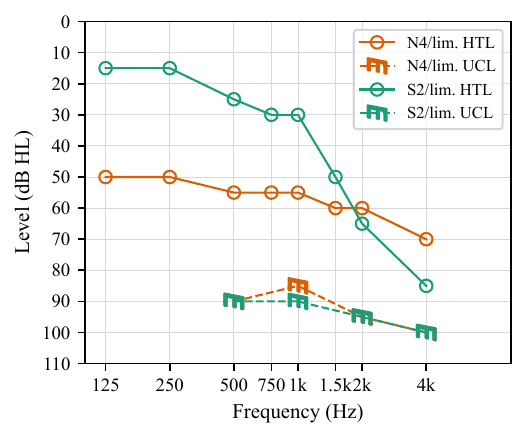}
    \caption{Audiograms N4 and S2 with limited residual DR used for optimization. Solid lines: hearing thresholds (HTL); scattered lines:
    individually measured UCLs.}
    \label{fig:audiograms}
\end{figure}
}
 
\subsection{Optimization Results}
 
Table~\ref{tab:results_hl} and Fig.~\ref{fig:scatter} summarize the optimization outcomes. All optimized WDRC configurations maintain full envelope-level UCL compliance (GUE\,$\geq$\,0.99) and near-perfect transient compliance (TGUE\,$\geq$\,0.98), unlike NAL-R which shows transient exceedances (TGUE\,=\,0.91--0.97).

{
\begin{table}[ht]
\caption{Optimization results for profiles N4 and S2 with limited residual DR, averaged over 55 and 90 dB SPL input levels. GR and G2 are prescription based and include an optimized broadband gain offset.}
\label{tab:results_hl}
\centering
\footnotesize
\setlength{\tabcolsep}{4pt}
 
\textbf{N4 with limited DR}
 
\vspace{2pt}
\begin{tabular}{@{}ll c cc cc cc@{}}
\toprule
& & \multirow{2}{*}{GR}
& \multicolumn{2}{c}{WDRC-1}
& \multicolumn{2}{c}{WDRC-7}
& \multicolumn{2}{c}{WDRC-20} \\
\cmidrule(lr){4-5} \cmidrule(lr){6-7} \cmidrule(lr){8-9}
& & & G2 & Opt. & G2 & Opt. & G2 & Opt. \\
\midrule
\multirow{3}{*}{HAAQI $\uparrow$}
& Tight   & 1.00 & 0.99 & 0.98 & 0.79 & 0.93 & 0.78 & 0.89 \\
& Normal  & 1.00 & 0.99 & 0.99 & 0.76 & 0.94 & 0.72 & 0.94 \\
& Dynamic & 1.00 & 0.92 & 0.87 & 0.71 & 0.82 & 0.73 & 0.83 \\
\midrule\midrule
\multirow{3}{*}{GLE $\uparrow$}
& Tight   & 0.90 & 0.89 & 0.95 & 0.93 & 0.96 & 0.95 & 0.93 \\
& Normal  & 0.92 & 0.91 & 0.95 & 0.95 & 0.96 & 0.96 & 0.95 \\
& Dynamic & 0.75 & 0.66 & 0.72 & 0.79 & 0.81 & 0.80 & 0.81 \\
\bottomrule
\end{tabular}
 
\vspace{6pt}
\textbf{S2 with limited DR}
 
\vspace{2pt}
\begin{tabular}{@{}ll c cc cc cc@{}}
\toprule
& & \multirow{2}{*}{GR}
& \multicolumn{2}{c}{WDRC-1}
& \multicolumn{2}{c}{WDRC-7}
& \multicolumn{2}{c}{WDRC-20} \\
\cmidrule(lr){4-5} \cmidrule(lr){6-7} \cmidrule(lr){8-9}
& & & G2 & Opt. & G2 & Opt. & G2 & Opt. \\
\midrule
\multirow{3}{*}{HAAQI $\uparrow$}
& Tight   & 1.00 & 0.98 & 0.97 & 0.79 & 0.93 & 0.79 & 0.92 \\
& Normal  & 1.00 & 0.98 & 0.97 & 0.83 & 0.94 & 0.84 & 0.93 \\
& Dynamic & 1.00 & 0.90 & 0.86 & 0.79 & 0.82 & 0.77 & 0.81 \\
\midrule\midrule
\multirow{3}{*}{GLE $\uparrow$}
& Tight   & 0.87 & 0.81 & 0.86 & 0.85 & 0.87 & 0.87 & 0.92 \\
& Normal  & 0.87 & 0.82 & 0.88 & 0.86 & 0.93 & 0.88 & 0.94 \\
& Dynamic & 0.81 & 0.72 & 0.77 & 0.77 & 0.83 & 0.79 & 0.84 \\
\bottomrule
\end{tabular}
\end{table}
}

{
\begin{figure}[!ht]
    \centering
    \includegraphics[width=\linewidth]{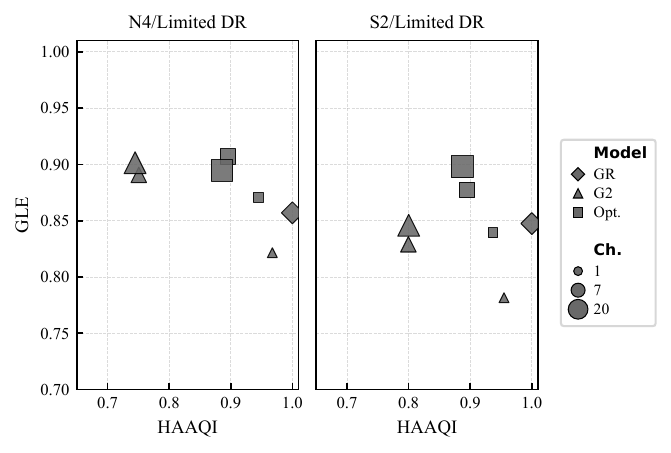}
    \caption{HAAQI vs.\ GLE for profiles N4 and S2 with limited residual DR. Symbol shape indicates model; symbol size indicates channel count. Values averaged over 55 and 90 dB SPL for the three signals.}
    \label{fig:scatter}
\end{figure}
}

The gain-optimized G2 baseline yields high HAAQI for WDRC-1 (0.90--0.99) but substantially lower scores for multi-band compressors (0.71--0.84), indicating that compression parameters fitted to NAL-NL2 speech targets introduce considerable distortion when applied to music. Per-signal optimization improves HAAQI by 0.12 on average (range 0.03--0.22). All quality improvement stems from compression parameter tuning: the gain offset does not affect HAAQI but accounts for most of the GLE improvement, with the gain fraction strongly audiogram-dependent (79\% for flat N4 vs.\ 50\% for sloping S2), consistent with the established quality--compression trade--off in fitting practice~\cite{kirchbergerDynamicRangeMusic2016a, croghanMusicPreferencesHearing2014a}.

A quality--audibility trade-off emerges across channel counts. From the optimized models, WDRC-1 achieves the highest HAAQI, exceeding multi-band models by 0.05 on average. However, WDRC-1 optimization reduces HAAQI compared to the single-channel G2 baseline (by up to 0.05), indicating that the single-channel optimizer sacrifices quality to improve audibility. This is a constraint that multi-band models resolve through frequency-specific compression. HAAQI also evaluates against a linear-gain reference, structurally favoring single-channel processing. The multi-band audibility advantage is audiogram-shape-dependent: for the flat N4 profile, WDRC-7 and WDRC-20 improve GLE over WDRC-1 only for the \textit{Dynamic} signal (+0.09), with negligible differences for \textit{Tight} and \textit{Normal}. For the sloping S2 profile, WDRC-20 achieves the highest GLE across all three signals (0.92, 0.94, 0.84), exceeding WDRC-1 by 0.06 on average, because the finer frequency resolution addresses the varying residual DR landscape more precisely. Fig.~\ref{fig:scatter} visualizes this trade-off: no single configuration dominates on both axes.
 
\section{Conclusion}
\label{sec:conclusion}

This paper investigated whether music fitting in hearing aids benefits from being signal-aware on two levels: assessing compression demand, and, where linear gain is insufficient, tuning compression parameters to the signal--audiogram combination.

(1) Music dynamics are an important determinant of how strongly compression is indicated, comparable in effect size to residual DR severity. For the mild-to-moderate profiles evaluated, production-compressed recordings fit the residual DR under linear gain alone, which is preferable whenever audibility is achieved without compression~\cite{kirchbergerDynamicRangeMusic2016a, croghanMusicPreferencesHearing2014a}. The analysis indicates the strongest case for compression in dynamically rich signals and/or limited residual DR, motivating signal--audiogram--adaptive decisions rather than fixed audiogram-based music-program configurations.

(2) Per signal--audiogram optimization improves HAAQI over gain-optimized NAL-NL2 (G2) target baselines while also improving audibility. Being maximally granular over the static compression characteristic establishes the achievable ceiling within this parameter space and quantifies the headroom available to less granular, deployable strategies. Gain adjustment and compression tuning serve largely orthogonal roles. The results indicate that the benefit of multi-band compression is audiogram-shape-dependent: the sloping profile gains from finer frequency resolution.

All results are based on objective metrics; subjective listening tests are the essential next step to validate the observed gains. Extending the optimization to attack and release times is expected to further improve the quality--audibility trade-off, and broader signal and audiometric coverage would support deployable signal-adaptive fitting strategies that approximate the ceiling established here.

\newpage

\bibliographystyle{IEEEbib}
\bibliography{refs}

@misc{akibaOptunaNextgenerationHyperparameter2019,
  title = {Optuna: {{A Next-generation Hyperparameter Optimization Framework}}},
  shorttitle = {Optuna},
  author = {Akiba, Takuya and Sano, Shotaro and Yanase, Toshihiko and Ohta, Takeru and Koyama, Masanori},
  year = 2019,
  month = jul,
  number = {arXiv:1907.10902},
  eprint = {1907.10902},
  primaryclass = {cs},
  publisher = {arXiv},
  doi = {10.48550/arXiv.1907.10902},
  urldate = {2026-04-01},
  archiveprefix = {arXiv}
}

@article{bisgaardStandardAudiogramsIEC2010a,
  title = {Standard {{Audiograms}} for the {{IEC}} 60118-15 {{Measurement Procedure}}},
  author = {Bisgaard, Nikolai and Vlaming, Marcel S. M. G. and Dahlquist, Martin},
  year = 2010,
  month = jun,
  journal = {Trends Amplif},
  volume = {14},
  number = {2},
  pages = {113--120},
  issn = {1084-7138},
  doi = {10.1177/1084713810379609},
  urldate = {2026-04-01},
  pmcid = {PMC4111352},
  pmid = {20724358}
}

@article{byrneNationalAcousticLaboratories1986,
  title = {The {{National Acoustic Laboratories}}' ({{NAL}}) {{New Procedure}} for {{Selecting}} the {{Gain}} and {{Frequency Response}} of a {{Hearing Aid}}},
  author = {Byrne, Denis and Dillon, Harvey},
  year = 1986,
  month = aug,
  journal = {Ear and Hearing},
  volume = {7},
  number = {4},
  pages = {257--265},
  langid = {english}
}

@article{croghanChosenListeningLevels2016,
  title = {Chosen {{Listening Levels}} for {{Music With}} and {{Without}} the {{Use}} of {{Hearing Aids}}},
  author = {Croghan, Naomi B. H. and Swanberg, Anne M. and Anderson, Melinda C. and Arehart, Kathryn H.},
  year = 2016,
  month = sep,
  journal = {Am J Audiol},
  volume = {25},
  number = {3},
  pages = {161--166},
  issn = {1059-0889, 1558-9137},
  doi = {10.1044/2016_AJA-15-0078},
  urldate = {2026-04-13},
  langid = {english}
}

@article{croghanMusicPreferencesHearing2014a,
  title = {Music {{Preferences With Hearing Aids}}: {{Effects}} of {{Signal Properties}}, {{Compression Settings}}, and {{Listener Characteristics}}},
  author = {Croghan, Naomi B H and Arehart, Kathryn H and Kates, James M},
  year = 2014,
  month = sep,
  journal = {Ear \& Hearing},
  langid = {english}
}

@misc{defferrardFMADatasetMusic2017,
  title = {{{FMA}}: {{A Dataset For Music Analysis}}},
  shorttitle = {{{FMA}}},
  author = {Defferrard, Micha{\"e}l and Benzi, Kirell and Vandergheynst, Pierre and Bresson, Xavier},
  year = 2017,
  month = sep,
  number = {arXiv:1612.01840},
  eprint = {1612.01840},
  primaryclass = {cs},
  publisher = {arXiv},
  doi = {10.48550/arXiv.1612.01840},
  urldate = {2026-04-01},
  archiveprefix = {arXiv}
}

@article{fortunetoddDurationCompressionAided2000,
  title = {Duration, {{Compression}}, and the {{Aided Loudness Discomfort Level}}},
  author = {{Fortune, Todd} and {Scheller, Tom}},
  year = 2000,
  month = aug,
  journal = {Ear and Hearing},
  volume = {21},
  number = {4},
  pages = {329--341}
}

@article{goncalvesbrazOPRARSHearingAidFitting2022,
  title = {{{OPRA-RS}}: {{A Hearing-Aid Fitting Method Based}} on {{Automatic Speech Recognition}} and {{Random Search}}},
  shorttitle = {{{OPRA-RS}}},
  author = {Gon{\c c}alves Braz, Libio and Fontan, Lionel and Pinquier, Julien and Stone, Michael A. and F{\"u}llgrabe, Christian},
  year = 2022,
  month = feb,
  journal = {Front. Neurosci.},
  volume = {16},
  pages = {779048},
  issn = {1662-453X},
  doi = {10.3389/fnins.2022.779048},
  urldate = {2026-04-21},
  langid = {english}
}

@article{greasleyMusicListeningHearing2020a,
  title = {Music Listening and Hearing Aids: Perspectives from Audiologists and Their Patients},
  shorttitle = {Music Listening and Hearing Aids},
  author = {Greasley, Alinka and Crook, Harriet and Fulford, Robert},
  year = 2020,
  month = sep,
  journal = {International Journal of Audiology},
  volume = {59},
  number = {9},
  pages = {694--706},
  issn = {1499-2027, 1708-8186},
  doi = {10.1080/14992027.2020.1762126},
  urldate = {2026-04-01},
  langid = {english}
}

@article{greasleyUsingHearingAids2026a,
  title = {Using {{Hearing Aids}} for {{Music}}: {{A UK Survey}} of {{Challenges}} and {{Strategies}}},
  shorttitle = {Using {{Hearing Aids}} for {{Music}}},
  author = {Greasley, Alinka E. and Beeston, Amy V. and Fulford, Robert J. and Crook, Harriet and Salter, Jackie M. and Hake, Robin and Moore, Brian C. J.},
  year = 2026,
  month = jan,
  journal = {Trends Hear},
  volume = {30},
  pages = {23312165251396517},
  issn = {2331-2165},
  doi = {10.1177/23312165251396517},
  urldate = {2026-03-31},
  pmcid = {PMC12833179},
  pmid = {41568998}
}

@article{jafriOldenburgHearingHealth2025,
  title = {The {{Oldenburg Hearing Health Record}} ({{OHHR}})},
  author = {Jafri, Sumbul and Berg, Daniel and Buhl, Mareike and Vormann, Matthias and Saak, Samira and Wagener, Kirsten C. and Thiel, Christiane M. and Hildebrandt, Andrea and Kollmeier, Birger},
  year = 2025,
  month = sep,
  journal = {Sci Data},
  volume = {12},
  number = {1},
  pages = {1546},
  issn = {2052-4463},
  doi = {10.1038/s41597-025-05884-y},
  urldate = {2026-04-01},
  langid = {english}
}

@article{katesHearingAidAudioQuality2016a,
  title = {The {{Hearing-Aid Audio Quality Index}} ({{HAAQI}})},
  author = {Kates, James M. and Arehart, Kathryn H.},
  year = 2016,
  month = feb,
  journal = {IEEE/ACM Trans Audio Speech Lang Process},
  volume = {24},
  number = {2},
  pages = {354--365},
  issn = {2329-9290},
  doi = {10.1109/TASLP.2015.2507858},
  urldate = {2026-03-31},
  pmcid = {PMC4849486},
  pmid = {27135042}
}

@article{keidserNALNL2PrescriptionProcedure2011,
  title = {The {{NAL-NL2 Prescription Procedure}}},
  author = {Keidser, G. and Dillon, H. and Flax, M. and Ching, T. and Brewer, S.},
  year = 2011,
  month = mar,
  journal = {Audiol Res},
  volume = {1},
  number = {1},
  issn = {2039-4330},
  doi = {10.4081/audiores.2011.e24},
  urldate = {2026-04-01},
  pmcid = {PMC4627149},
  pmid = {26557309}
}

@article{kirchbergerDynamicRangeMusic2016a,
  title = {Dynamic {{Range Across Music Genres}} and the {{Perception}} of {{Dynamic Compression}} in {{Hearing-Impaired Listeners}}},
  author = {Kirchberger, Martin and Russo, Frank A.},
  year = 2016,
  month = jan,
  journal = {Trends in Hearing},
  volume = {20},
  pages = {2331216516630549},
  issn = {2331-2165, 2331-2165},
  doi = {10.1177/2331216516630549},
  urldate = {2026-03-31},
  langid = {english}
}

@misc{loshchilovCMAESHyperparameterOptimization2016,
  title = {{{CMA-ES}} for {{Hyperparameter Optimization}} of {{Deep Neural Networks}}},
  author = {Loshchilov, Ilya and Hutter, Frank},
  year = 2016,
  month = apr,
  number = {arXiv:1604.07269},
  eprint = {1604.07269},
  primaryclass = {cs},
  publisher = {arXiv},
  doi = {10.48550/arXiv.1604.07269},
  urldate = {2026-04-01},
  archiveprefix = {arXiv}
}

@book{mooreCochlearHearingLoss2007a,
  title = {Cochlear {{Hearing Loss}}: {{Physiological}}, {{Psychological}} and {{Technical Issues}}},
  shorttitle = {Cochlear {{Hearing Loss}}},
  author = {Moore, Brian C. J.},
  year = 2007,
  month = jan,
  edition = {2nd ed},
  publisher = {John Wiley \& Sons},
  address = {Chichester Hoboken, N.J},
  isbn = {978-0-470-51818-2 978-0-470-51633-1 978-0-470-98788-9},
  langid = {english}
}

@article{mooreEffectsSoundInducedHearing2016,
  title = {Effects of {{Sound-Induced Hearing Loss}} and {{Hearing Aids}} on the {{Perception}} of {{Music}}},
  author = {Moore, Brian},
  year = 2016,
  month = mar,
  journal = {J. Audio Eng. Soc.},
  volume = {64},
  number = {3},
  pages = {112--123},
  issn = {15494950},
  doi = {10.17743/jaes.2015.0081},
  urldate = {2026-04-13},
  langid = {english}
}

@inproceedings{pascoedavidpedroClinicalMeasurementsAuditory1988,
  title = {Clinical Measurements of the Auditory Dynamic Range and Their Relation to Formulas for Hearing Aid Gain},
  booktitle = {Hearing Aid Fitting: {{Theoretical}} and Practical Views},
  author = {{Pascoe, David Pedro}},
  year = 1988,
  volume = {13},
  pages = {129--152}
}

@article{serpanosEffectSoundPreference2025,
  title = {Effect of {{Sound Preference}} on {{Loudness Tolerance}} and {{Preferred Listening Levels Using Personal Listening Devices}}},
  author = {Serpanos, Yula C. and DiBlasi, Thomas and Butler, Jasmin},
  year = 2025,
  month = jun,
  journal = {Audiology Research},
  volume = {15},
  number = {3},
  pages = {68},
  issn = {2039-4349},
  doi = {10.3390/audiolres15030068},
  urldate = {2026-04-13},
  langid = {english}
}

@incollection{Zakis2016,
  title = {Music Perception and Hearing Aids},
  booktitle = {Hearing Aids},
  author = {Zakis, Justin A.},
  editor = {Popelka, Gerald R. and Moore, Brian C. J. and Fay, Richard R. and Popper, Arthur N.},
  year = 2016,
  month = sep,
  pages = {217--252},
  publisher = {Springer International Publishing},
  address = {Cham},
  doi = {10.1007/978-3-319-33036-5_8},
  isbn = {978-3-319-33036-5}
}

@article{zwislockiTemporalSummationLoudness1969,
  title = {Temporal {{Summation}} of {{Loudness}}: {{An Analysis}}},
  shorttitle = {Temporal {{Summation}} of {{Loudness}}},
  author = {Zwislocki, J. J.},
  year = 1969,
  month = aug,
  journal = {The Journal of the Acoustical Society of America},
  volume = {46},
  number = {2B},
  pages = {431--441},
  issn = {0001-4966, 1520-8524},
  doi = {10.1121/1.1911708},
  urldate = {2026-04-01},
  langid = {english}
}
 
\end{document}